\documentclass[12pt]{article}
\usepackage[dvips]{color}
\usepackage{epsfig}
\usepackage{amsmath}
\usepackage{graphicx}

\begin{document}

\title{\bf The phase transition and holographic in modified Horava -Lifshitz black hole}

\author{{Kh. Jafarzade $^{a}$ \thanks{Email: kh.Jafarzadeh@stu.umz.ac.ir} \hspace{1mm},
J. Sadeghi $^{a}$\thanks{Email:
pouriya@ipm.ir}} \\
$^a${\small {\em Sciences Faculty, Department of Physics, University
of Mazandaran,}}\\{\small {\em P. O. Box 47415-416, Babolsar,
Iran}}}

\maketitle

\begin{abstract}
In this paper, we take cosmological constant as a thermodynamical
pressure and it's conjugate quantity as a thermodynamical volume.
This expression help us to investigate the phase transition and
holographic heat engine. So, in order to have Van der Waals fluid
behaviour in Horava -Lifshitz (HL) black hole, we modified the
solution of such black hole with some cosmology ansatz.
Also from  holographic heat engine, we obtain Carnot efficiency for the HL black hole. The phase transition of the system lead us to investigate the stability condition for the corresponding black hole. In that case, we show that the stability exist only in special region of black hole.\\

\end{abstract}
\section{Introduction}
Bekenstain, Hawking and Bardeen compared the laws of black hole
mechanics with the ordinary thermodynamics. They used the
appropriate quantities as temperature, entropy and energy and etc
and clarified all laws of black hole. In that case, they have shown
that all laws of black hole mechanics are identical to the
laws of the ordinary thermodynamics and hence the black hole is considered as a thermal system [1-2].\\
In recent years, the thermodynamic properties and critical behaviour
of various black holes investigated by Ref.s. [3-7]. But this topic
will be attractive when we study black hole thermodynamic in the
presence of a negative cosmological constant. The most important
example is AdS black holes with various phase transitions. The
physics of such black holes has been improved by the AdS/CFT
correspondence. The primary study phase transition of AdS black hole is 
started by Hawking-Page phase transition and it demonstrated phase
transition between the Schwarzschild AdS black hole and thermal AdS
space [8]. On the other hand, one of the interesting results is
investigation of phase transition Reissner-Nordstrom-AdS (RN-AdS)
black hole. The study of charged AdS black holes indicated that they
have behaviour like as liquid-gas system. In this case, the
cosmological constant and it's conjugate quantity treat as
thermodynamic pressure and thermodynamic volume respectively.\\
The influence of different topologies on the $P-V$ criticality of
black hole can not be ignored in the extended phase space [6,9-11].
In this regard some efforts have been made such as study of
thermodynamic of topological black hole solutions in Einstein
gravity [12,13], Einstein-Gauss- Bonnet gravity, dilaton gravity
[14-16] and Lovelock gravity [17]. The investigation of charged AdS
black holes in Einstein gravity indicated that the critical behavior
exists only for a black hole with horizon of spherical topology. But
for planner and hyperbolic cases there is not any critical behavior.
Similar investigation has been made for charged black hole in Gauss-
Bonnet gravity and the obtained results was similar to charged AdS
black holes in Einstein gravity. The research of black hole
solutions in Horava-Lifshitz gravity will be interesting and also is different
to Einstein  gravity theory [18-23].  Because,  the phase transition
of charged (uncharged) topological black holes in Horava-Lifshitz
gravity only take place in $k=-1$ (hyperbolic horizon) in the
nonextended phase space. This consequence does not satisfy in the
extended phase space. In this case, for all horizon topologies there
is not any physical critical point for the uncharged and charged
cases. Therefore, topological black hole in Horava-Lifshitz gravity
like  BTZ black hole, have not any phase transition in the extended
phase space. So, by looking to the two black holes as BTZ and
Horava-Lifshitz, one can say that the existence of cosmological
constant in the black hole metric is not sufficient for the
existence $P-V$ criticality. Therefore, the
problem of phase transition in HL black hole give us motivation to
modify the solution of such system  with suitable ansatz. In that
case it's
thermodynamic quantities match exactly with Van der Waals fluid equation state.\\
Here, we   study the P-V criticality in classical mechanic.  It may be interesting to  investigate such behaviour in quantum  mechanic . From quantum approach, one can
consider effect of thermal fluctuations where they interpret as
quantum effect. Thermal fluctuations arise due to quantum
fluctuations in the geometry of space-time and they appear in the
black hole entropy as logarithmic term [24-27]. But it's not our
purpose in this paper. Here, we
study thermodynamic black hole from the classical point of view.\\
So, we are going to recall some classical thermodynamic relations
which are necessary
for the investigation of thermodynamic and phase transition of black hole.\\
The quantities which constitute base of black hole thermodynamics
are as follows,
\begin{equation}
T=\frac{\kappa\hbar}{2\pi
k_{B}c}~~~~,~~~~S_{H}=\frac{A}{4L^{2}_{PL}}~~~~,~~~~E=Mc^{2},
\end{equation}
where $\kappa$, $L_{PL}=(\frac{\hbar G}{c^{3}})^{\frac{1}{2}}$, $ A$
and $E$ are the surface gravity, the Planck length, area of event
horizon and energy (mass) of black hole respectively. One can extend
this classical subject with definition of pressure and volume. The
cosmological constant is associated by thermodynamic pressure as
[6,7],
\begin{equation}
P=-\frac{\Lambda}{8\pi},
\end{equation}
and it's thermodynamic conjugate of pressure is interpreted as a
black hole thermodynamic volume,
\begin{equation}
V=\bigg( \frac{\partial M}{\partial P}\bigg)_{S,...}.
\end{equation}
The cosmological constant can be changed in the first law of black
hole thermodynamic, so in this case the first law in the extended
phase space will be as,
\begin{equation}
dM=TdS+VdP+...~.
\end{equation}
In extended phase space, the mass is not related to the internal
energy $U$ of the system, but it is  related by enthalpy which is
given by,
\begin{equation}
M=H\equiv U+PV.
\end{equation}
Here we note that,  when $P$ is constant (the cosmological constant is not allowed to
vary), the relation (4)  reduces to the standard first law of
thermodynamics in the "non-extended" phase space.

By using relation (2) and (3), we can write black hole equation of
state $P=P(V,T)$ and compare to the corresponding fluid. In that case, we identify the black hole and fluid quantities
as $T\sim T_{f}$, $V\sim V_{f}$  and $P\sim P_{f}$ [28].\\
On the other hand, the Van der Waals equation of state is
modification of the ideal gas which includes the nonzero size and
attraction between corresponding molecules. It is used to describe
basic qualitative features of the liquid-gas phase transition,
\begin{equation}
kT=(P+\frac{a}{\upsilon^{2}})(\upsilon -b),
\end{equation}
here $\upsilon = V/N$ is the specific volume of the fluid, $P$, $T$
and $k$ are pressure, temperature and the Boltzmann constant
respectively. The constant $b>0$ indicate the nonzero size of the
molecules of fluid and the attraction between them is indicated by
$a>0$. After studying of phase transition, it is interesting to
define classical cycles for black holes like usual thermodynamic
systems. It means that when small black hole translate to large
black hole we need again the large black hole translate to small
black hole, in other words the system must be back to primary state.
This definition interpret as holographic heat engine which is mentioned for first
time by [29-31]. We are going to investigate this subject at the last section of this paper. So, we organize the corresponding paper as follows. In
Sect.2, we will review the solution of black hole in Horava-Lifshitz
gravity. In Sect.3, In order to have phase transition we will modify
Lu-Mei-Pop (LMP) solution of HL black hole. In that case, we show
that the corresponding stress energy tensor does not obey any of the
energy conditions everywhere outside of the horizon, but the weak
condition will be satisfied in near horizon. Then we investigate the
stability of solution and determine stable and unstable phase at the
end of this section. In Sect.4, we investigate holography stuff and
exhibit suitable cycle for corresponding heat engine. Also we
arrange special condition to have Carnot efficiency. Finally, in
section 5 we have some results and conclusion.
\section{Horava-Lifshitz black hole solution}
One of the interesting kinds of black holes is the Horava-Lifshitz,
which may be regarded as a $UV$ candidate for general relativity.
The HL is a non-relativistic renormalizable theory of gravity at a
Lifshitz point. It provides an interesting framework to study the
connection between ordinary gravity and string theory. It is
expected that the HL black hole solutions asymptotically become
Einstein gravity solutions. Also it is found that a Schwarzschild
AdS solution can be realized in
infrared modified HL gravity theory [23,32,33].\\
The four-dimensional gravity action of HL theory is given by the
following expression [34-36],
\begin{eqnarray}
S_{HL}=\int
dtd^{3}x\sqrt{g}N\bigg(\frac{2}{\kappa^{2}}(K_{ij}K^{ij}-\lambda
K^{2})+\frac{\kappa^{2}\mu^{2}(\Lambda_{W}R-3\Lambda_{W}^{2})}{8(1-3\lambda)}+\frac{\kappa^{2}\mu^{2}(1-4\lambda)}{32(1-3\lambda)}R^{2}\bigg)\nonumber
\\&& \hspace{-155mm}-\int
dtd^{3}x\sqrt{g}N\bigg(\frac{\kappa^{2}\mu^{2}}{8}R_{ij}R^{ij}+\frac{\kappa^{2}\mu}{2\omega^{2}}\epsilon^{ijk}R_{il}\nabla_{j}R^{l}_{k}-\frac{\kappa^{2}}{2\omega^{4}}C_{ij}C^{ij}\bigg),
\end{eqnarray}
where $\kappa^{2}, \Lambda_{W},\omega,\lambda$ and $\mu$ are
constant parameters and $C_{ij}$ is Cotton tensor,
\begin{equation}
C^{ij}=\epsilon^{ikl}\nabla_{k}\bigg(R^{j}_{l}-\frac{1}{4}R\delta^{j}_{l}\bigg),
\end{equation}
and $K_{ij}$ is the extrinsic curvature,
\begin{equation}
K_{ij}=\frac{1}{2N}(g_{ij}-\nabla_{i}N_{j}-\nabla_{j}N_{i}),
\end{equation}
where $N_{i}$ and $N$ are shift and lapse functions respectively.
Also the cosmological constant is given by the following relation,
\begin{equation}
\Lambda=\frac{3}{2}\Lambda_{W}.
\end{equation}
The HL black hole metric is,
\begin{equation}
ds^{2}=-f(r)dt^{2}+\frac{dr^{2}}{f(r)}+r^{2}d\Omega^{2},
\end{equation}
where $f (r)$ is given by [37-39],
\begin{equation}
f(r)=k+(\omega-\Lambda_{W})r^{2}-\sqrt{(r(\omega(\omega-2\Lambda_{W})r^{3}+\beta))},
\end{equation}
and $\beta$ is integration constant. There are two limits of HL
black hole which is following. The first one is Kehagius-Sfetsos
(KS) solution which is obtained by $\Lambda_{W}=0$  and
$\beta=4\omega M$ [40]. In that case the equation (12) reduces to
the following expression,
\begin{equation}
f_{ks}(r)=k+\omega r^{2}-\omega r^{2}\sqrt{1+\frac{4M}{\omega
r^{3}}}.
\end{equation}
The second one is Lu-Mei-Pop (LMP) solution  which is obtained by
$\omega=0$ and $\beta=-\frac{\alpha^{2}}{\Lambda_{W}}$, where
$\alpha$ related to the black hole mass $(\alpha = aM)$ [41]. In
that case the equation (12) reduces to the following expression,
\begin{equation}
f_{LMP}(r)=k-\Lambda_{W}r^{2}-\alpha\sqrt{\frac{r}{-\Lambda_{W}}}.
\end{equation}
The charged topological black holes in Horava-Lifshitz gravity were
investigated by Ref.s [22,23]. When the dynamical coupling constant
$\lambda$ is set to one, the corresponding metric is given by,
\begin{equation}
f(r)=k+x^{2}-\sqrt{c_{0} x-\frac{q^{2}}{2}},
\end{equation}
where $x=\sqrt{-\Lambda}r$ and
$\Lambda=(-\frac{3}{l^{2}})$ corresponds to the negative cosmological
constant. The physical mass and the charge ($Q$) corresponding to the
black hole solution are respectively given by,
\begin{equation}
M=\frac{\kappa^{2}\mu^{2}\Omega_{k}\sqrt{-\Lambda}}{16}c_{0}~~~;~~~Q=\frac{\kappa^{2}\mu^{2}\Omega_{k}\sqrt{-\Lambda}}{16}q,
\end{equation}
where $c_{0}$ and $q$ are the integration constants, $\Omega_{k}$ is
the volume of the two dimensional Einstein space and $\kappa$ and
$\mu$ are the constant parameters of the theory. The event horizon
can be obtained by $f(r_{+}) = 0$ where for Eq.(15) is as follows,
\begin{equation}
x^{4}+Ax^{2}+Bx+c=0,
\end{equation}
where the coefficients are as: $A=2k$, $B=-c_{0}$ and
$C=(k^{2}+\frac{q^{2}}{2})$. Applying the Descarte method [42] in
equation (17), we find the following solution,
\begin{eqnarray}
x_{1}=\frac{-a-\sqrt{-(a^{2}+2A+\frac{2B}{a})}}{2},~~~x_{2}=\frac{-a+\sqrt{-(a^{2}+2A+\frac{2B}{a})}}{2},\nonumber
\\&&
\hspace{-125mm}x_{3}=\frac{a-\sqrt{-(a^{2}+2A-\frac{2B}{a})}}{2},~~~x_{4}=\frac{a+\sqrt{-(a^{2}+2A-\frac{2B}{a})}}{2},
\end{eqnarray}
with
\begin{eqnarray}
a=\pm\bigg[\bigg(-\frac{q}{2}-\frac{1}{2}\sqrt{\frac{27q^{2}+4p^{3}}{27}}
\bigg)^{\frac{1}{3}}+
\bigg(-\frac{q}{2}+\frac{1}{2}\sqrt{\frac{27q^{2}+4p^{3}}{27}}-\frac{2A}{3}
\bigg)^{\frac{1}{3}}\bigg]^{\frac{1}{2}}\nonumber
\\&&\hspace{-138mm}b=\frac{1}{2}\bigg(a^{2}+A+\frac{B}{a}\bigg),~~~~~~~~~~~~~c=\frac{1}{2}\bigg(a^{2}+A-\frac{B}{a}\bigg)\nonumber
\\&&\hspace{-138mm}p=-4C-\frac{A^{2}}{3},~~~~~~~~~~~~~~~~~~~~q=\frac{27A}{3}\bigg(A^{2}+36C\bigg)+C.
\end{eqnarray}
Here the horizon corresponds to a positive choice of solutions,
\begin{equation}
x_{h}=\frac{a+\sqrt{-(a^{2}+2A-\frac{2B}{a})}}{2}.
\end{equation}
The black hole entropy, Hawking temperature and heat capacity at
constant pressure determined by thermodynamical relation [22,43,44],
\begin{equation}
S=\int\frac{dM}{T}~~~,~~~T_{H}=\frac{\kappa}{2\pi}=\frac{1}{4\pi}\bigg(\frac{\partial
f}{\partial r}\bigg)_{r=r_{h}}~~~,~~~C_{P}=\frac{\partial
M}{\partial T}=\frac{T}{(\frac{\partial T}{\partial S})_{P}}.
\end{equation}
By using these relations, one can obtain thermodynamical quantities
of (LMP) solution Eq.(14) as follows,
\begin{equation}
S=\frac{8\pi}{a}\sqrt{-\Lambda_{W}r_{h}}~~,~~T_{H}=\frac{1}{8\pi
r_{h}}(-k-3\Lambda_{W}r_{h}^{2})~~,~~C_{P}=\frac{4\pi}{a}\sqrt{-\Lambda_{W}r_{h}}\frac{k+3\Lambda_{W}r_{h}^{2}}{-k+3\Lambda_{W}r_{h}^{2}}.
\end{equation}
The heat capacity at constant pressure has a discontinuity at
$r_{h}=\sqrt{\frac{k}{3\Lambda_{W}}}$. This shows that there is a
second-order phase transition at this point. The cosmological
constant is negative in AdS space, therefore phase transition exits
only in $k=-1$. This is completely different to Einstein theory,
where we have phase transition for only $k=1$ case. Also
thermodynamical quantities for solution (15) are given by,
\begin{eqnarray}
T_{H}=\frac{\sqrt{-\Lambda}(3x_{h}^{4}+2kx_{h}^{2}-k^{2}-\frac{Q^{2}}{2})}{8\pi
x_{h}(k+x_{h}^{2})},\nonumber
\\&&\hspace{-72mm}S=\frac{x_{h}^{2}}{4}+\frac{k}{2}\ln
x_{h}+S_{0},\nonumber
\\&&\hspace{-72mm}C_{P}=\frac{(k+x_{h}^{2})^{2}(3x_{h}^{4}+2kx_{h}^{2}-k^{2}-\frac{Q^{2}}{2})}{6x_{h}^{6}+14kx_{h}^{4}+10k^{2}x_{h}^{2}+2k^{3}+Q^{2}(k+3x_{h}^{2})}.
\end{eqnarray}
The heat capacity suffers a discontinuity at $x_{h}=x_{c}$ in
$k=-1$. Then this solution has a phase transition in nonextended
phase space for hyperbolic horizon. The critical behavior of charged
topological black holes in Horava-Lifshitz gravity in the extended
phase space was studied in [23] and no physical critical point was
found in all the cases $(k=\pm1,0)$.
\section{Modification of  Horava-Lifshitz black hole solution}
 Here the most important thing is investigation of the thermodynamical and
hydrodynamical properties
 Horava-Lifshitz black hole. Also, the phase transition and  Van der Waals fluid behaviour for such black hole play important role in holography and AdS/CFT. So, for this reason we first investigated the  usual Horava-Lifshitz black hole solution and observed that there are not such properties. Therefor we take advantage from some ansatz and energy condition and modify the  usual  Horava-Lifshitz black hole to new form of solution which is modification of  Horava-Lifshitz black hole solution. In that case we see this new form of HL are satisfied by phase transition and  Van der Waals fluid  properties. So, in this section, we are going to  modify the equation (15) for the
uncharge case, where thermodynamics properties of black hole exactly
coincide with the given fluid equation of state. In order to modify
such metric background, first we give following ansatz for the
$f(r)$,
\begin{eqnarray}
ds^{2}=-f(r)dt^{2}+\frac{dr^{2}}{f(r)}+r^{2}d\Omega^{2}\nonumber
\\&& \hspace{-67mm} f(r)=-\Lambda r^{2}-2\sqrt{Mr}-h(r,P).
\end{eqnarray}
In order to simplify the corresponding results,  we  set $\frac{\kappa^{2}\mu^{2}\Omega_{k}}{4}=1$.

Now we are going to determine the function $h(r,P)$ which has to be
satisfied with the equation of motion. So, generally we assume that
the above metric is a solution of the HL gravity with given energy
momentum source from HL lagrangian. In order to the energy-momentum
source be physically acceptable
 it should be satisfied by certain following conditions:\\
1)The energy density be positive.\\
 2)The energy density conquest to
pressure.\\
These conditions are known as energy conditions, e.g. [45], which is
given by
\begin{eqnarray}
Week: ~~~\rho\geq0~~~~,~~~\rho+p_{i}\geq0\nonumber \\&&
\hspace{-65mm}
Strong:~~~\rho+\sum_{i}p_{i}\geq0~~~,~~~\rho+p_{i}\geq0\nonumber
\\&& \hspace{-65mm}
Dominant:~~~\rho\geq\mid p_{i}\mid,
\end{eqnarray}
where $\rho$ and $p_{i}$ are  energy density and
principal pressure respectively. Also, $\rho$ and $p_{i}$ clarified by the  following relations,
\begin{eqnarray}
\rho=-p_{1}=\frac{1-f-r f'}{8\pi r^{2}}+P,\nonumber \\&&
\hspace{-59mm} p_{2}=p_{3}=\frac{r f''+2f'}{16\pi r}-P.
\end{eqnarray}
After  determining $f(r)$, we examine the corresponding energy
conditions. The  mass  of the black hole  $(M)$ is related to the
horizon radius $r_{+}$ according to the following expression,
\begin{equation}
\sqrt{M}=\frac{1}{2\sqrt{r_{+}}}(-\Lambda r_{+}^{2}-h(r_{+},P)).
\end{equation}
On the other hand the black hole temperature  is given by the following equation,
\begin{equation}
T=\frac{1}{4\pi}f'(r_{+})=3Pr_{+}+
\frac{h(r_{+},P)}{8Pr_{+}}-\frac{h'(r_{+},P)}{4\pi}.
\end{equation}

As a mentioned above, we have  to construct a metric which exactly
matches thermodynamics properties of black hole to the thermodynamic
of Van der Waals fluid. For this purpose, we compare the temperature
of $T$ in equation (28) with (6) and use $\upsilon=3r_{+}$,
\begin{equation}
3Pr_{+}+
\frac{h(r_{+},P)}{8Pr_{+}}-\frac{h'(r_{+},P)}{4\pi}-P\upsilon+b\upsilon-\frac{a}{\upsilon}+\frac{ab}{\upsilon^{2}}=0,
\end{equation}
Here we use the following ansatz [28,46],
\begin{equation}
h(r,P)=A(r)-PB(r).
\end{equation}
By substituting the equation (30) into  equation (29), $B(r)$ and
$A(r)$ will be as,
\begin{equation}
B(r)=-8b\pi r+C_{1}\sqrt{r}.
\end{equation}
and
\begin{equation}
A(r)=\frac{8a\pi}{3}-\frac{8ab\pi}{27r}+C_{2}\sqrt{r}.
\end{equation}
Here $C_{1}$ and $C_{2}$ are integration constants and also $C_{2}$
complectly has a mass dimension and $C_{1}$  is  mass over pressure.
By replacing  $A(r)$ and $B(r)$ in the relation (24) we have,
\begin{equation}
f(r)=-\frac{8a\pi}{3}+\frac{8ab\pi}{27r}-C\sqrt{Mr}+\Lambda
br-\Lambda r^{2},
\end{equation}
where $C$ is a constant in terms of $C_{1}$ and $C_{2}$. Since $a$
is positive, we conclude that the topology of horizon is hyperbolic.
Without loss of generality we can set $a=\frac{3}{8\pi}$. We can
determine event horizon by using $f(r_{+}=0)$ and draw its
variations with respect to $b$ and $C$ parameters. As we see in
figure (1), the event horizon increases with $b$ parameter linearly
but it decreases by increasing $C$ and will be zero for very large
values of $C$.

\begin{figure}
\hspace*{1cm}
\begin{center}
\epsfig{file=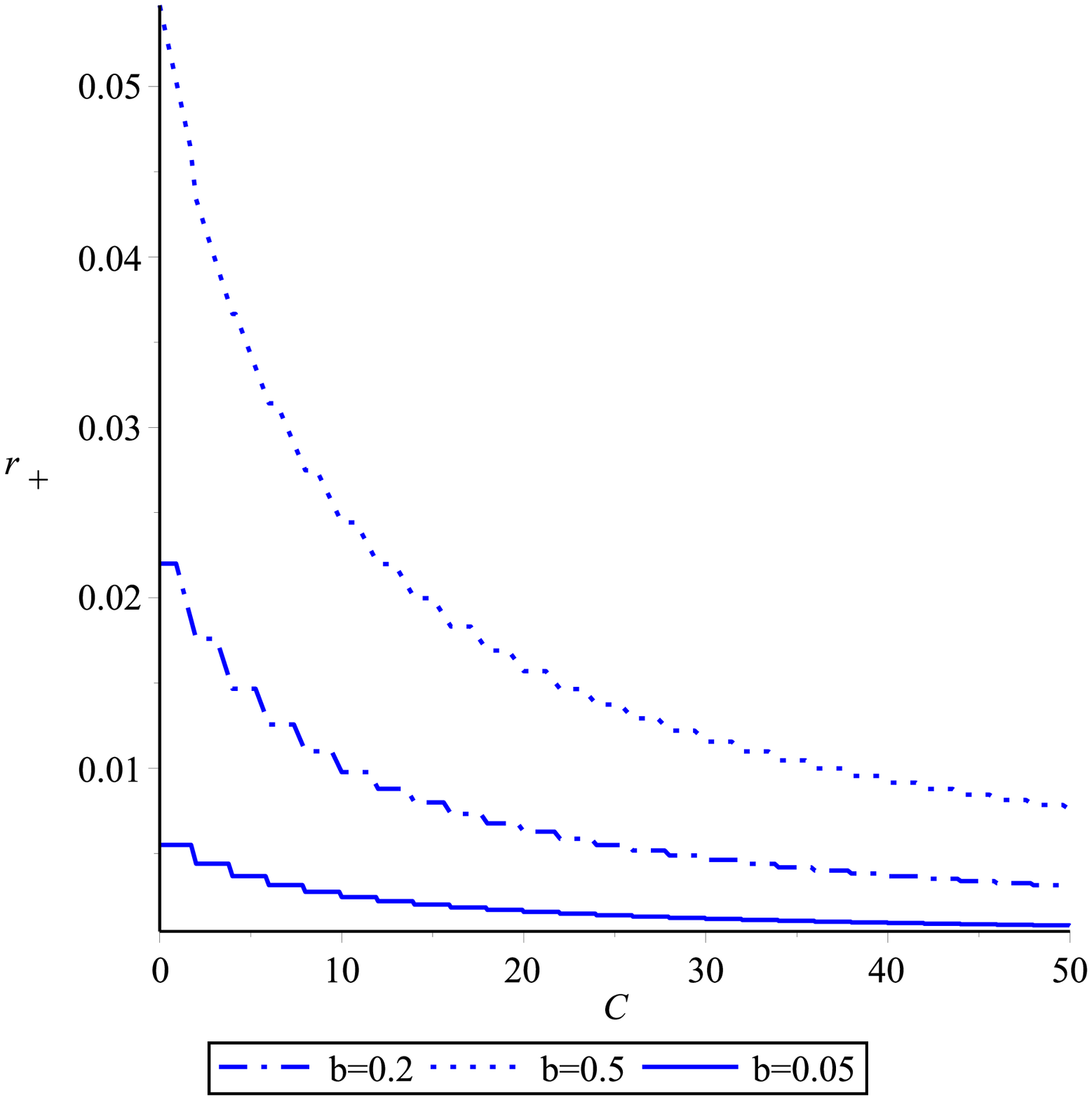,width=7cm}
\epsfig{file=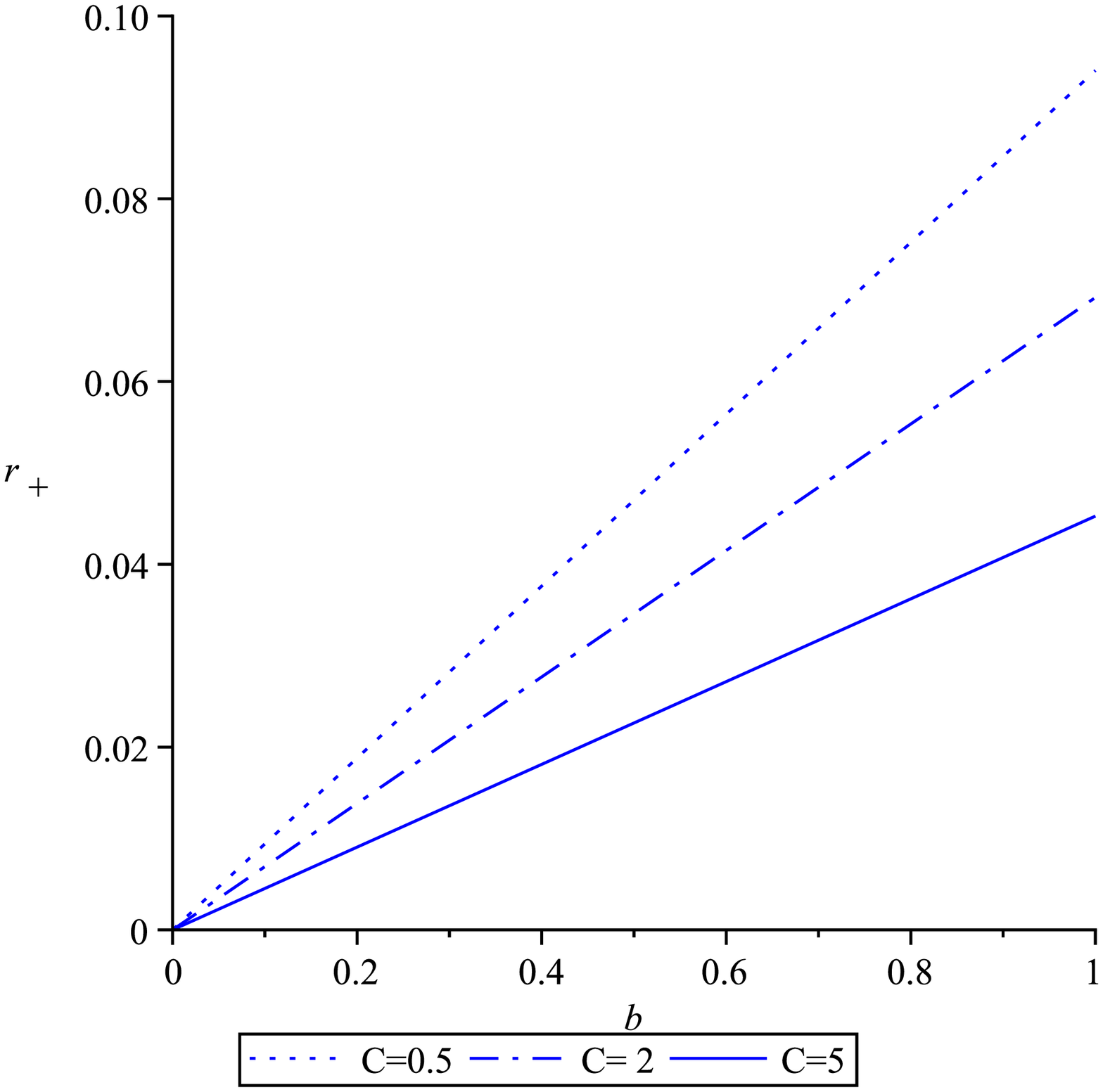,width=7cm}\caption{\small{The variation of event
horizon with respect to $b$ parameter for $P=0.01$ and $M=0.4$ with
different $C$ (right plot) and the variation of event horizon with
respect to $C$ parameter for $P=0.01$ and $M=0.4$ with different $b$
(left plot).}}
\end{center}
\end{figure}

Now by rearranging $f(r)$, we can investigate the energy condition.
So, by using relation (26), we obtain $\rho$ and $p_{i}$,
\begin{eqnarray}
\rho=-p_{1}=\frac{1}{16\pi r}(\frac{b}{3r^{2}}+\frac{1}{r}+8b\pi
P-8\pi Pr)\nonumber
\\&&
 \hspace{-85mm}
p_{2}=p_{3}=\frac{1}{16\pi r}(-\frac{b}{12r^{2}}+\frac{3}{4r}-10b\pi
P+26\pi Pr)\nonumber
\\&&
 \hspace{-85mm}\rho+p_{2}=\frac{1}{16\pi r}(\frac{b}{4r^{2}}+\frac{7}{4r}-2b\pi
P+18\pi Pr).
\end{eqnarray}

We display the energy density $\rho$ (Solid and Dash dot curves) and
the quantity $\rho+P_{2}$ (Dot curve) with respect to $r_{+}$. We
always have $\rho+P_{1}=0$, but for the arbitrary pressure and
everywhere outside of the horizon, we have $\rho+P_{2}\geq0$.  Also
note that  if the pressure be sufficiently small ($P=0.001$),  the
energy condition $\rho\geq0$ will be  satisfy near  horizon,  which
is illustrated by  Fig (2).
\begin{figure}
\hspace*{1cm}
\begin{center}
\epsfig{file=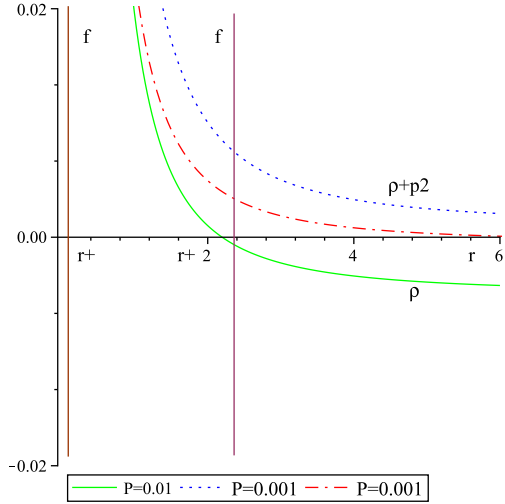,width=7cm} \caption{\small{The variation of energy condition with respect to horizon. In this figure we  set $M=0.4$, $b=0.5$ and
$C=0.4$.}}
\end{center}
\end{figure}
In HL black hole with comparing to thermodynamical system we have
some phase transition. In order to find the stability and
instability phase transition we have to investigate the
corresponding subject. As we know the stability of system from phase
transition point of view play important role in some phenomena in
particle physics and also cosmology system. So, for this reasons we
consider stability condition for the corresponding system. In that
case, we need two quantities as Gibbs free energy and heat capacity
which play important role for the study of stability system. When
the Gibbs free energy is negative $(G<0)$, the system has a global
stability. For the local stability equilibrium in thermodynamic
system requires that $C_{P}\geq C_{V}\geq0$ [47,48]. In that case,
the heat capacity at constant pressure and constant volume is given
by,
\begin{equation}
C_{V}=\frac{\partial H}{\partial T}\bigg|_V=T\frac{\partial
S}{\partial T}\bigg|_V~~~~;~~~~C_{P}=\frac{\partial H}{\partial
T}P=T\frac{\partial S}{\partial T}\bigg|_P.
\end{equation}
The temperature and the entropy can be obtained by equation (21),
which are related by the first law of thermodynamics.
\begin{equation}
T=\frac{1}{4\pi}f'(r_{+})=\frac{1}{4\pi}(\frac{1}{2r_{+}}-\frac{b}{6r_{+}^{2}}-4b\pi
P+12\pi Pr_{+}).
\end{equation}
We emphasis that the  area formula for entropy  breaks down in case
of higher derivative gravity especially HL theory [43]. For this
reason we leave this formula and employ the following first law of
thermodynamics and obtain the corresponding entropy as,
\begin{equation}
S=\int\frac{dM}{T}=\frac{8\pi}{C^{2}}(-\ln(r_{+})-\frac{b}{9r_{+}}-8b\pi
Pr_{+}+4\pi Pr_{+}^{2})+S_{0}.
\end{equation}
In the above equation, $S_{0}$ is the integration constant. By using
Eqs. (2) and (3), the thermodynamic volume can be calculated by,
\begin{equation}
V=\frac{16\pi}{C^{2}}(r_{+}-b)(-1+\frac{b}{9r_{+}}-8b\pi Pr_{+}+8\pi
Pr_{+}^{2}).
\end{equation}
As we see in equation (38) the volume depend on $r_{+}$ and $P$.
Then one can say, the heat capacity at constant volume vanishes,
\begin{equation}
C_{V}=T\bigg(\frac{\partial S}{\partial P}\frac{\partial P}{\partial
T}+\frac{\partial S}{\partial r_{+}}\frac{\partial r_{+}}{\partial
T}\bigg)=0.
\end{equation}
One can obtain $C_{P}$ as,
\begin{equation}
C_{P}=\frac{8\pi r_{+}}{C^{2}}\frac{(1-\frac{b}{3r_{+}}-8b\pi
Pr_{+}+24\pi Pr_{+}^{2})(-\frac{1}{r_{+}}+\frac{b}{9r_{+}^{2}}-8b\pi
P+8\pi Pr_{+})}{(-1+\frac{2b}{3r_{+}}+24\pi Pr_{+}^{2})}.
\end{equation}
In figure (3), We draw $C_{P}$ with respect to $r_{+}$ for two
values of $b$ parameter ($b=0.01$ and $b=0.5$).As we know, if the
heat capacity is divergent then the system will has a phase
transition. We notice that the heat capacity dose not diverge for
small values of $C$ then there is not the second-order phase
transition. As we see the heat capacity has a discontinuity for
small values of $b$.
 But by increasing this parameter ($b$) the
discontinuity points increases which can be decreased by increasing
$C$. For phase 1, the heat capacity is always positive which means
that this phase is thermodynamically stable. On the other hand  for
phase 2 the heat capacity is negative therefor it is an unstable
phase.

\begin{figure}
\hspace*{1cm}
\begin{center}
\epsfig{file=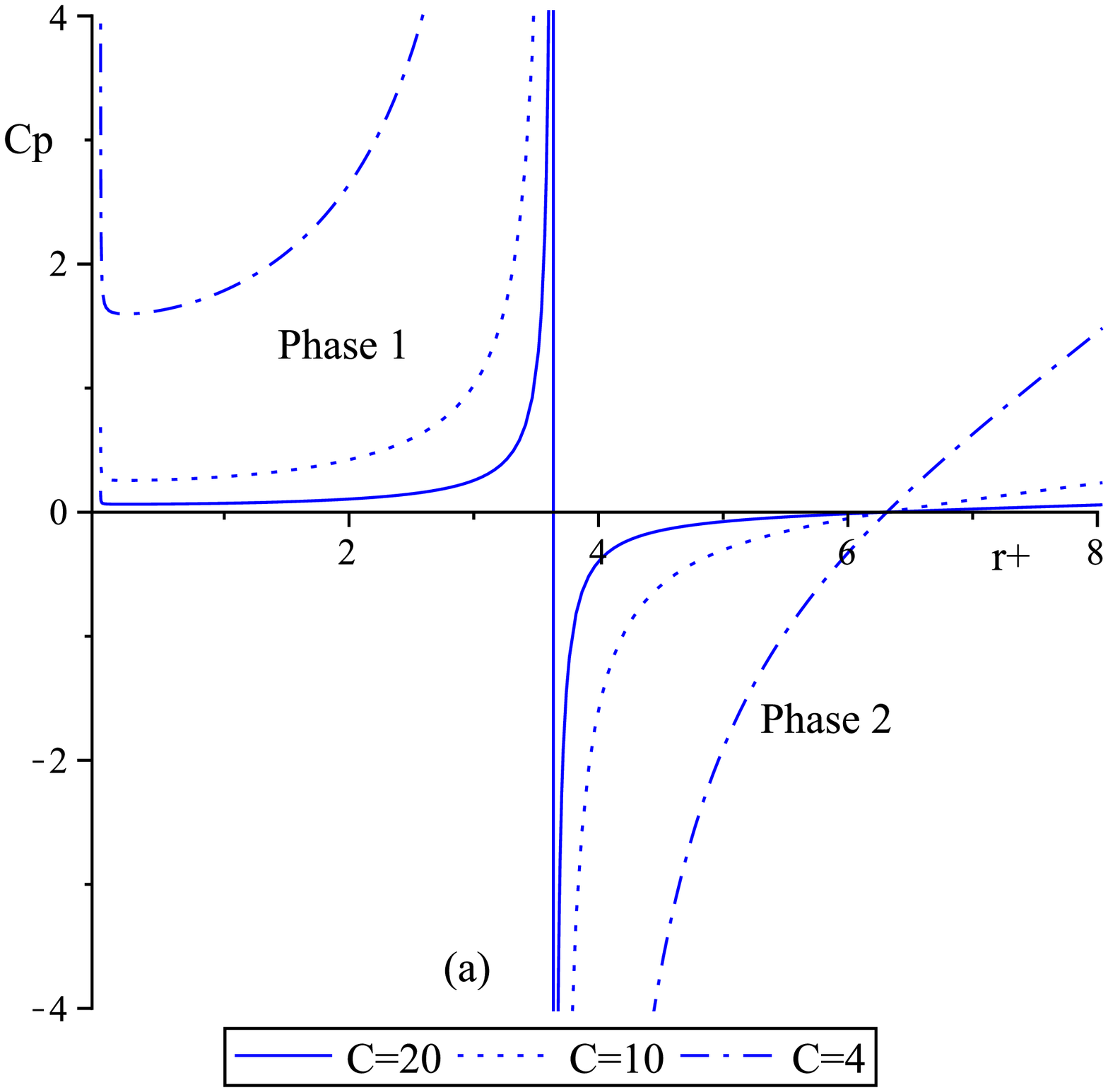,width=7cm}
\epsfig{file=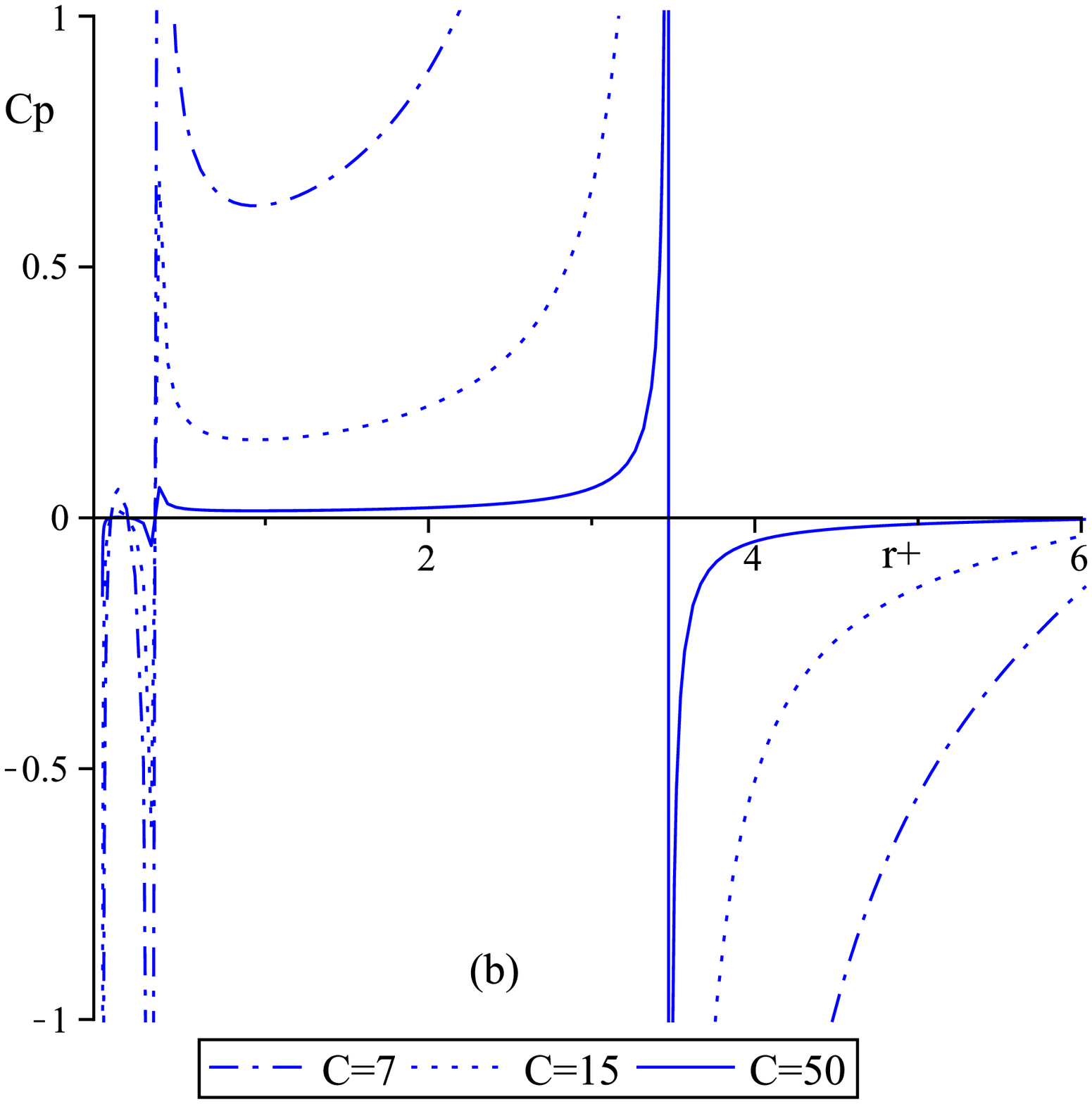,width=7cm}\caption{\small{(a) The variation of
heat capacity at constant pressure with respect to $r_{+}$ for
p=0.001, b=0.01. (b) The variation of  heat capacity at constant
pressure with respect to $r_{+}$ for p=0.001, b=0.5 with different
$C$.}}
\end{center}
\end{figure}

In order to discuss the global stability of the black hole,  we need to calculate the Gibbs free energy which  is given by,
\begin{equation}
G=E-TS+PV,
\end{equation}
and
\begin{equation}
G=\frac{1}{C^{2}r_{+}}(1-\frac{b}{9r_{+}}A_{1}-4\pi r_{+}^{2}A_{2}
)+\frac{64\pi^{2}}{C^{2}}Tr_{+}\ln(r_{+}),
\end{equation}
where
\begin{eqnarray}
A_{1}=1+\frac{2b}{9r_{+}}+48b\pi Pr_{+}-268\pi Pr_{+}^{2}\nonumber
\\&&
 \hspace{-74mm}
A_{2}=5-24b\pi P r_{+} +8\pi Pr_{+}^{2}.
\end{eqnarray}

\begin{figure}
\hspace*{1cm}
\begin{center}
\epsfig{file=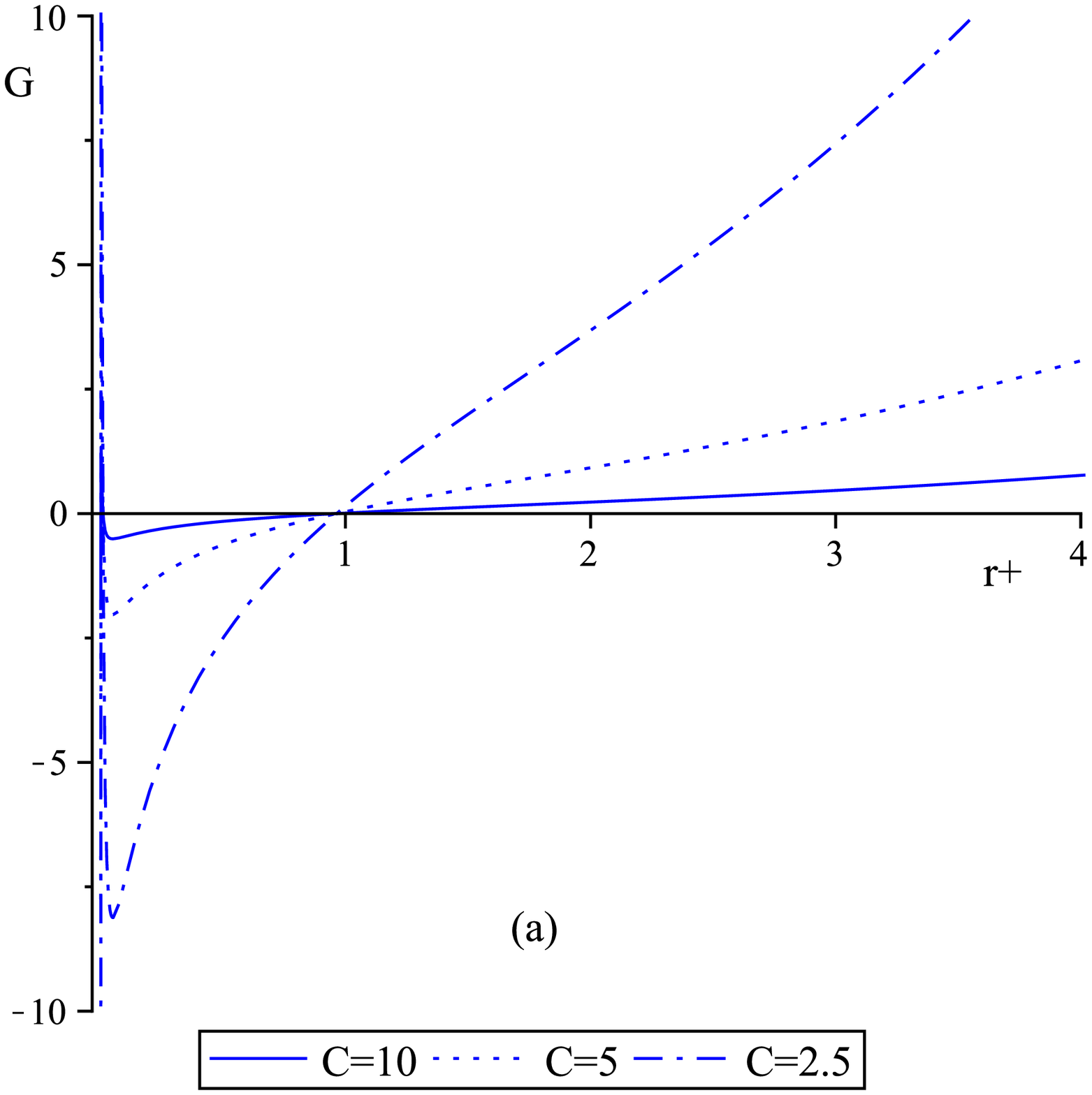,width=7cm}
\epsfig{file=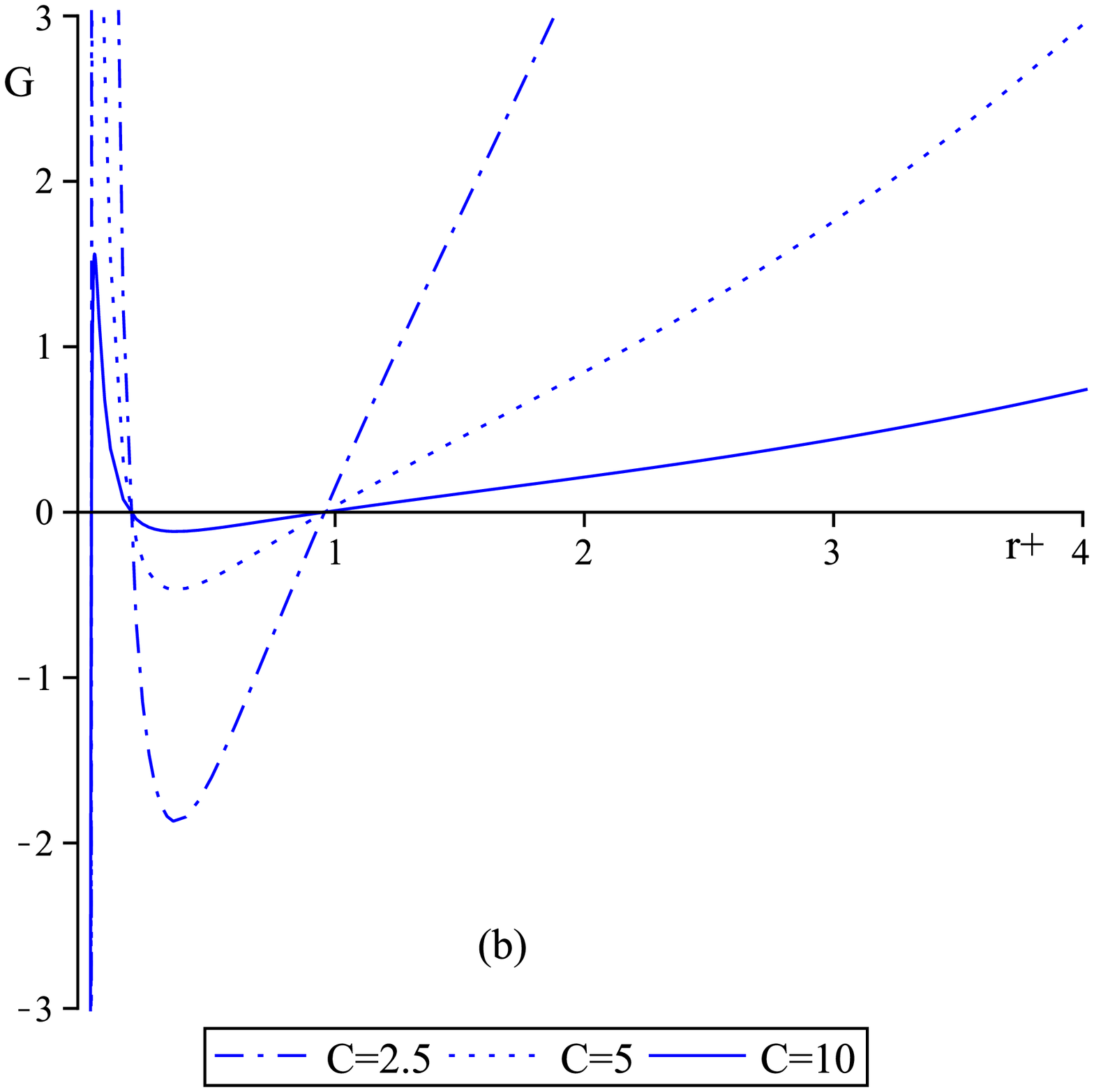,width=7cm}\caption{\small{(a) The variation of Gibbs free
energy  with respect to $r_{+}$ for  b=0.01, p=0.001
 and different values of $C$. (b) The variation of  Gibbs free
energy  with respect to $r_{+}$ for  b=0.5, p=0.001
 and different values of $C$.}}
\end{center}
\end{figure}
In figure (4), we display the Gibbs free energy with respect to
$r_{+}$. As we see in figure,  for large values of $C$ the system
has an instability. Here, we note that the range of stability
increases by decreasing $C$ and $b$.

\section{The corresponding holographic heat engine}
In this section, we intend to consider holographic heat engine for
the black hole solution. We start with equation of state of black
hole $P(V,T)$ which is relation between temperature, horizon radius,
other parameter of black hole as a cosmological constant. By using
the $P(V,T)$ function one can define an engine as a closed path in
$P-V$ plane. This lead us to have input and output heat as  $Q_{H}$
and $Q_{C}$ respectively. The total mechanical work  is
$W=Q_{H}-Q_{C}$ and also the efficiency of the heat engine is
determined by relation $\eta=\frac{W}{Q_H}=1-\frac{Q_C}{Q_H}$.\\
The properties of the heat engine depend on the equation of state
and choice of path in $P-V$ plane. For a heat engine which works
between two temperatures $T_{H}$ and $T_{C}$ the maximum efficiency will be Carnot efficiency [29-31].\\
The Carnot cycle is made from two isothermal and two adiabatic
processes. During isothermal expansion (constant temperature
$T_{H}$) the system absorbs heat of $Q_{H}$ and during isothermal
compression (constant
temperature $T_{C}$) loses heat of $Q_{C}$.\\
The  corresponding work done in this cycle is,
\begin{equation}
W=\oint PdV=\oint TdS=(T_{H}-T_{C})(S_{2}-S_{1}).
\end{equation}
And total amount of absorbed thermal energy will be as,
\begin{equation}
Q_{H}=T_{H}(S_{2}-S_{1}).
\end{equation}
The efficiency $\eta$ is defined by,
\begin{equation}
\eta=\frac{W}{Q_H}=1-\frac{T_C}{T_H}.
\end{equation}
This relation show that  the higher than  Carnot efficiency
 violate the second law of thermodynamics. Of course there is other method
to connect two isotherms $T_{H}$ and $T_{C}$ which are isochoric
paths. The formed cycle is called the Stirling
cycle, which has an efficiency less than the Carnot efficiency.\\
For static black holes the Carnot cycle and the Stirling cycle are
identical. Because in such black holes the entropy and  volume
both depend on the  horizon radius, so isochores and adiabatic are the same.\\

Now we study the efficiency of heat engine for modified solution and
investigate under what conditions, our cycle has the Carnot
efficiency. As we noticed in pervious section, the heat capacity at
constant volume vanishes for our solution. This is useful because it
can define another cycle in $P-V$ plane, as a rectangle which is
composed of isobars and isochores, see fig (5).
\begin{figure}
\hspace*{1cm}
\begin{center}
\epsfig{file=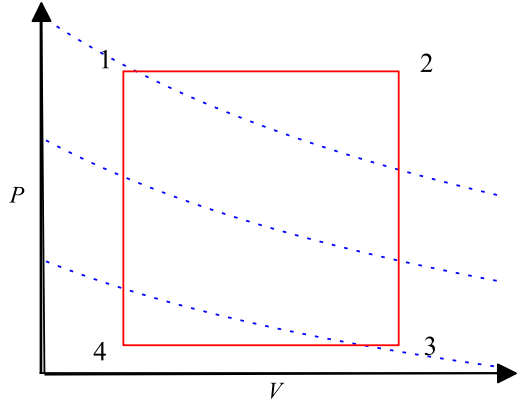,width=7cm} \caption{\small{The Carnot engine.}}
\end{center}
\end{figure}
The work done along the isobars is,
\begin{equation}
W=(V_{2}-V_{1})(P_{1}-P_{4}).
\end{equation}

 \begin{figure}
\hspace*{1cm}
\begin{center}
\epsfig{file=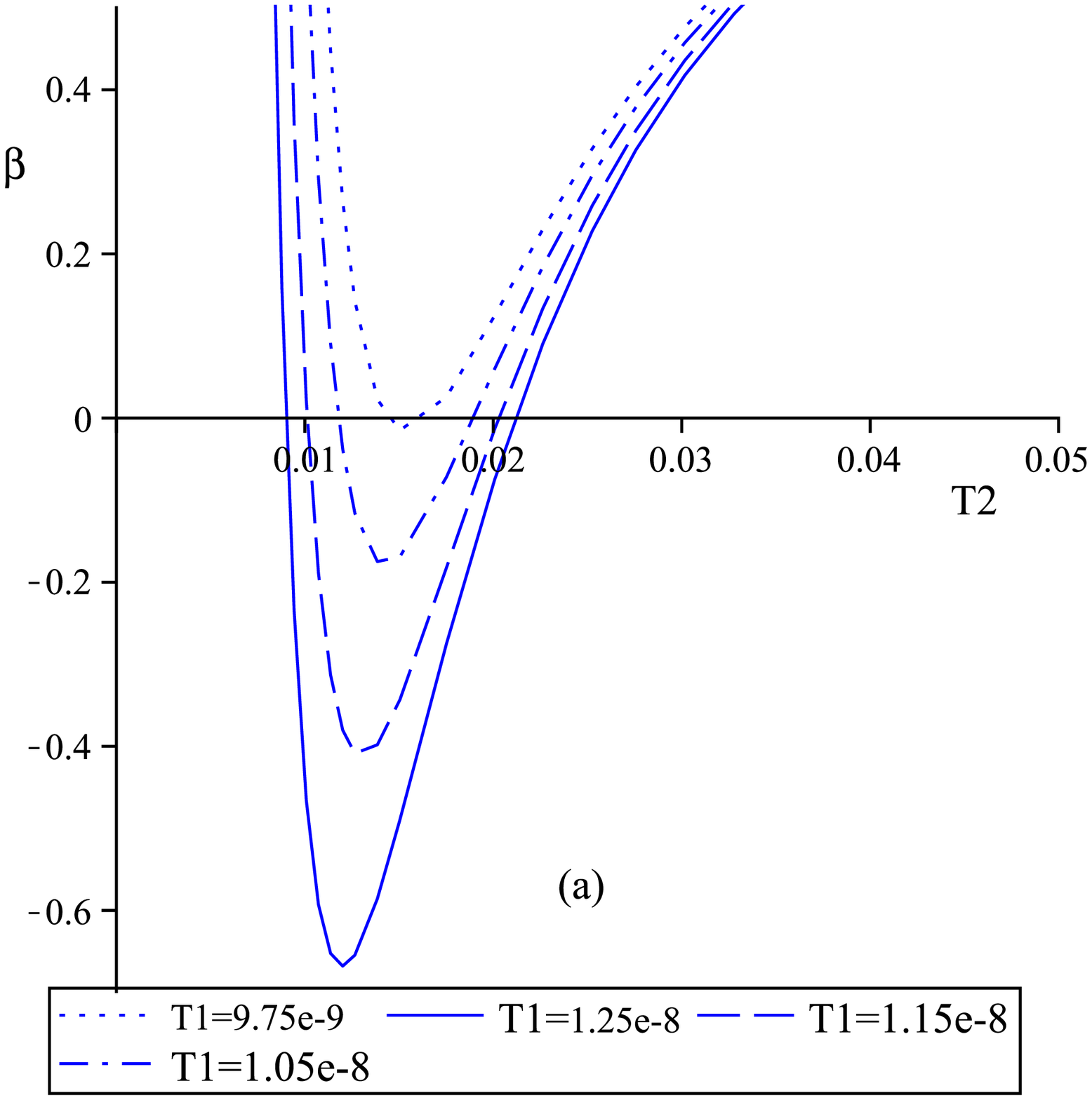,width=7cm}
\epsfig{file=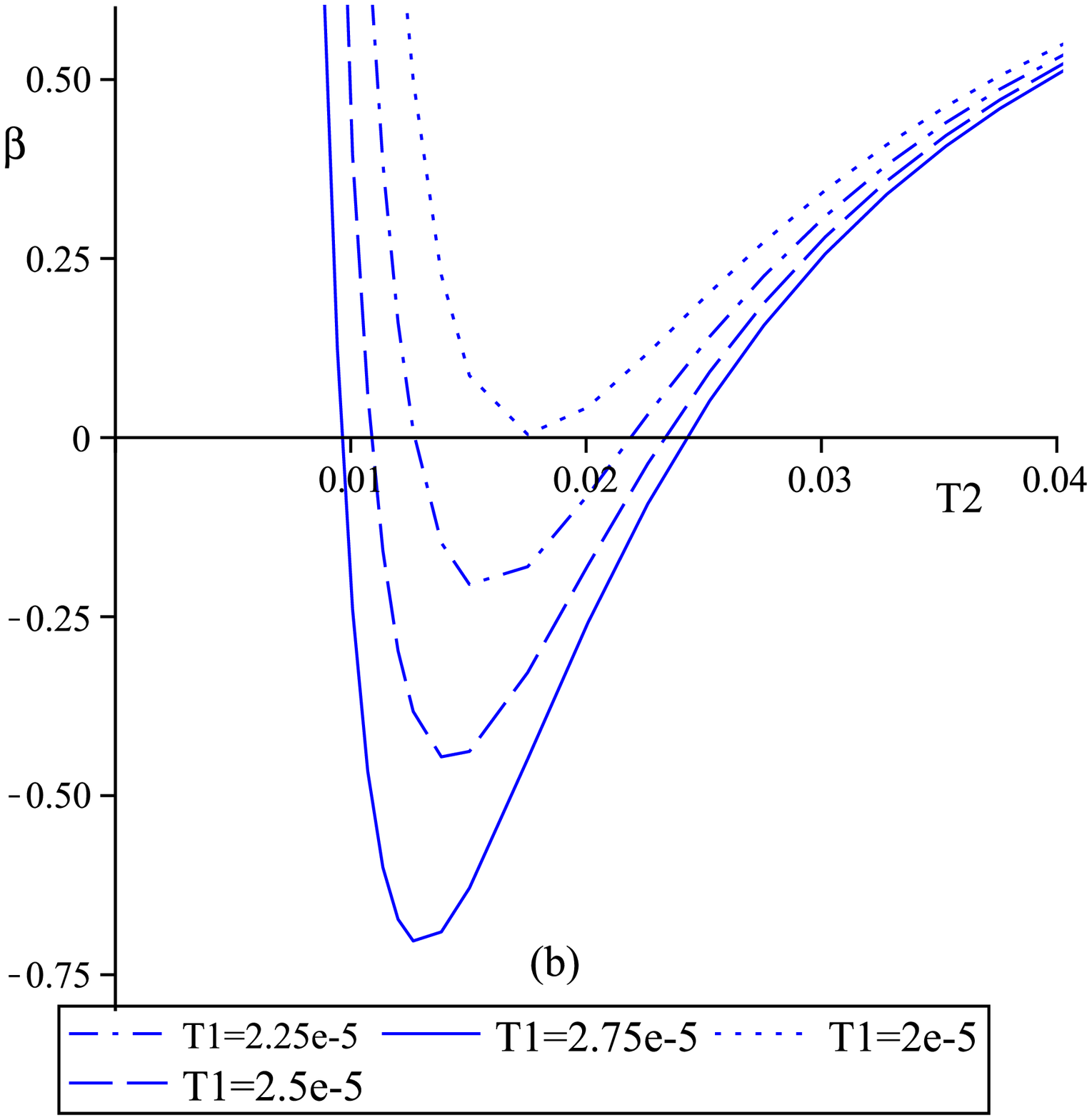,width=7cm}\caption{\small{(a) The variation of $\beta$  with
respect to temperature $T_{2}$ for P=0.001, b=0.01 and different
values of $T_{1}$. (b) The variation of $\beta$  with respect to temperature
$T_{2}$ for P=0.001, b=0.5 and different values of $T_{1}$.}}
\end{center}
\end{figure}

\begin{figure}
\hspace*{1cm}
\begin{center}
\epsfig{file=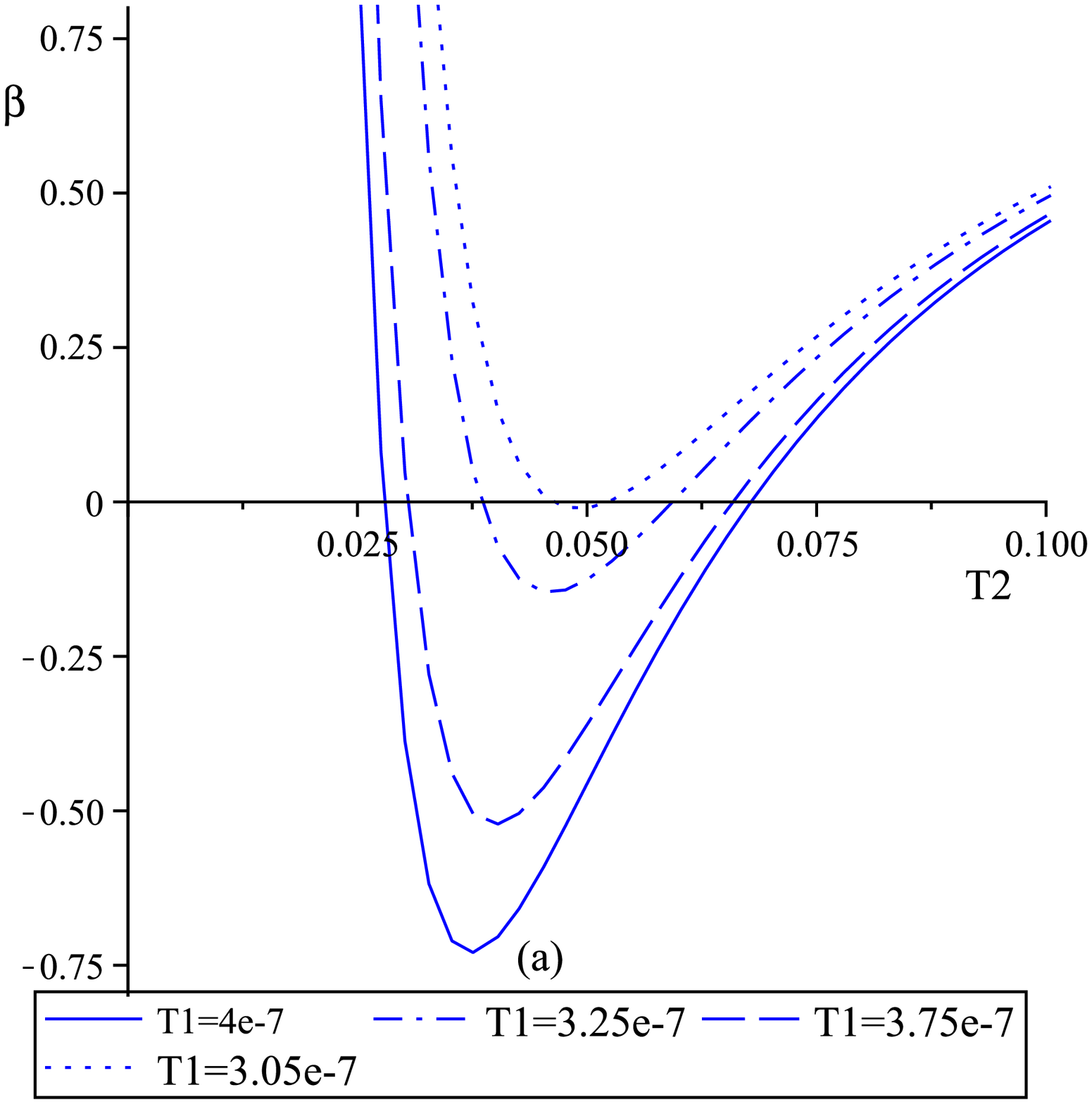,width=7cm}
\epsfig{file=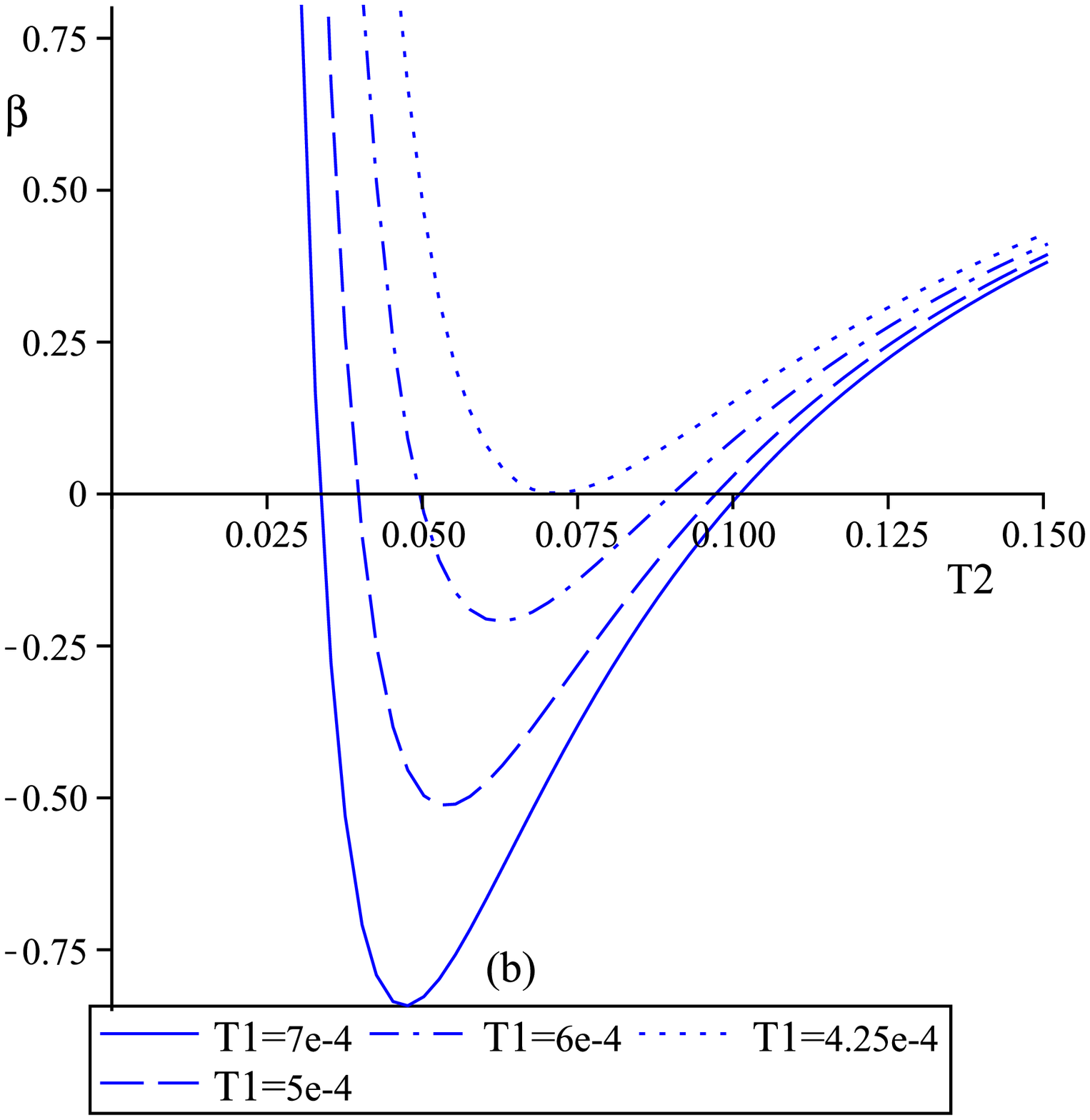,width=7cm}\caption{\small{(a) The variation of $\beta$  with
respect to temperature $T_{2}$ for P=0.01, b=0.01 and different
values of $T_{1}$. (b) The variation of $\beta$  with respect to temperature
$T_{2}$ for P=0.01, b=0.5 and different values of $T_{1}$.}}
\end{center}
\end{figure}

The heat flows occur along to the top and bottom of the figure.
Here, one can determine the inflow of heat with the upper isobar,
which is $Q_{H}$ ,
\begin{equation}
Q_H=\int_{T_1}^{T_2} C_P(P_1,T)dT.
\end{equation}
In order to calculate the integral, we need to write the heat
capacity in terms of $T$. So, this is complicated integral and one
can not solve directly because both $C_{P}$ and $T$ are  function of
$r_{+}$. In that case, we take a high temperature limit for
simplicity of calculation. In this limit we have,
\begin{equation}
r_{+}\sim \frac{T}{3P},
\end{equation}
and
\begin{equation}
C_P\sim \frac{64\pi^{2}}{9PC^{2}}T^{2},
\end{equation}
then
\begin{equation}
Q_H= \frac{64\pi^{2}}{27PC^{2}}(T_{2}^{3}-T_{1}^{3}).
\end{equation}
Also in this limit the thermodynamical volume is,
\begin{equation}
V= \frac{16\pi}{C^{2}}\bigg(\frac{8\pi T^{3}}{27P^{2}}-\frac{16b\pi
T^{2}}{9P}+\frac{(8b^{2}\pi P-1)T}{3P}-\frac{b^{2}P}{3T}\bigg).
\end{equation}

We substitute equation (52)  in equation (47) we have,

\begin{equation}
W=\frac{16\pi}{C^{2}}(P_{1}-P_{4})\bigg(\frac{8\pi}{27P^{2}}(T_{2}^{3}-T_{1}^{3})-\frac{16b\pi}{9P}(T_{2}^{2}-T_{1}^{2})+\frac{(8b^{2}\pi
P-1)}{3P}(T_{2}-T_{1})-\frac{b^{2}P}{3}(\frac{1}{T_{2}}-\frac{1}{T_{1}})\bigg).
\end{equation}
Therefore, the efficiency of this heat engine is,

\begin{equation}
\eta=\eta_{C}(1+\beta),
\end{equation}
where $\beta$,
\begin{equation}
\beta=1-12b\pi\frac{(T_{2}^{2}-T_{1}^{2})}{(T_{2}^{3}-T_{1}^{3})}+\frac{9P(8b^{2}\pi
P-1)}{4\pi}\frac{(T_{2}-T_{1})}{(T_{2}^{3}-T_{1}^{3})}-\frac{9b^{2}P^{3}}{4\pi}\frac{(T_{2}^{-1}-T_{1}^{-1})}{(T_{2}^{3}-T_{1}^{3})}~.
\end{equation}

Eq. (55) is satisfied by the condition $-1<\beta\leq0$. And if
$\beta=0$  we will have the carnot efficiency. Figures (6) and (7)
show the behavior of $\beta$ with respect to temperature $T_{2}$ for
different values of $T_{1}$. From figures (6) and (7), we see that
the system has efficiency in a certain range of $T_{1}$. And the
efficiency increases by reducing $T_{1}$ in this corresponding
certain range. Also the system will have the Carnot efficiency for
two specific values of $T_{2}$ . As we see from figure (6), by
increasing parameter $b$,  the $T_{1}$  also increases. By comparing
figures (6) and (7), we notice that by increasing the pressure, both
temperatures $T_{1}$ and $T_{2}$ are increasing.

\section{Conclusion}
As we know the investigation of  phase structure of Horava-Lifshitz
black hole has shown that the phase transition exists only for
hyperbolic topology of the horizon in the non-extended phase space.
This result shown that Horava-Lifshitz theory is not exactly
identical to Einstein theory. Because in Einstein theory exists no
phase transition for k=0, -1 but exist in case of  k=1. As we have seen, with
considering phase structure of HL black hole in the extended phase
space, there is  not any phase transition. For this reason\\
in this paper, we modified the  solution of the Horava-Lifshitz
black hole for uncharged case. We have shown that the modified
solution dose not satisfy any of three standard energy condition
everywhere outside of horizon. But for small pressure in near
horizon the modified solution satisfied by weak energy condition. We
investigated the stability and instability with certain
thermodynamical parameters. We noticed for small values of $b$ and
$C$ parameters the corresponding system has a global stability,  but
to have a phase transition the $C$ parameter must be large. Also we
saw that the divergence points of heat capacity increase by increasing
$b$ and $C$ parameters. Finally, we studied the heat engine and
observed that the system has a Carnot efficiency at special conditions.
For small values of pressure and $b$ parameter the system had a
maximum efficiency in range of very small $T_{1}$ , but this
temperature increased by increasing $b$ parameter. We also noticed
by increasing pressure the range of temperature $T_{2}$  increased.

\end{document}